# Spectrum Sensing Algorithms for Cognitive Radio Based on Statistical Covariances *

Yonghong Zeng, *Senior Member, IEEE* and Ying-Chang Liang, *Senior Member, IEEE*†

## Abstract

Spectrum sensing, i.e., detecting the presence of primary users in a licensed spectrum, is a fundamental problem in cognitive radio. Since the statistical covariances of received signal and noise are usually different, they can be used to differentiate the case where the primary user's signal is present from the case where there is only noise. In this paper, spectrum sensing algorithms are proposed based on the sample covariance matrix calculated from a limited number of received signal samples. Two test statistics are then extracted from the sample covariance matrix. A decision on the signal presence is made by comparing the two test statistics. Theoretical analysis for the proposed algorithms is given. Detection probability and associated threshold are found based on statistical theory. The methods do not need any information of the signal, the channel and noise power *a priori*. Also, no synchronization is needed. Simulations based on narrowband signals, captured digital television (DTV) signals and multiple antenna signals are presented to verify the methods.

## 1 Introduction

Conventional fixed spectrum allocation policy leads to low spectrum usage in many of the frequency bands. Cognitive radio, first proposed in [1], is a promising technology to exploit the under-utilized spectrum in an opportunistic manner [2, 3, 4, 5]. One application of cognitive radio is spectral reuse, which allows secondary networks/users to use the spectrum allocated/licensed to the primary users when they are not active [6]. To do so, the secondary users are required to frequently perform spectrum sensing, i.e., detecting the presence of the primary users. If the primary users are detected to be inactive, the secondary users can use the spectrum for communications. On the other hand, whenever the primary users become active, the secondary users have to detect the presence of those users in high probability, and vacate the channel within certain amount of time. One communication system using the spectrum reuse concept is IEEE 802.22 wireless regional area networks (WRAN) [7], which operates on the VHF/UHF bands that are currently allocated for TV broadcasting services and other services such as wireless microphone. Cognitive radio is also an emerging technology for vehicular devices. For example, in [8] cognitive radio is proposed for underwater vehicles to fully use the limited underwater acoustic bandwidth, and in [9] it is used for autonomous vehicular communications.

Spectrum sensing is a fundamental task for cognitive radio. However, there are several factors which make spectrum sensing practically challenging. First, the signal to noise ratio (SNR) of the primary users may be very low. For example, the wireless microphones operating in TV bands only transmit signals with a power of about 50mW and a bandwidth of 200 kHz. If the secondary users are several hundred meters away from the microphone devices, the received SNR may be well below $-20$dB. Secondly, multipath fading and time dispersion of the wireless channels make the sensing problem more difficult. Multipath fading may cause the signal power fluctuates as large as $20-30$dB. On the other hand, coherent detection may not be possible when the time dispersed channel is unknown, especially when the primary users are legacy systems which do not cooperate with the secondary users. Thirdly, the noise/interference level may change with time, which yields noise uncertainty. There are two types of noise uncertainty: receiver device noise uncertainty and environment noise uncertainty. The receiver device noise uncertainty comes from [10, 11, 12]: (a) non-linearity of components; and (b) time-varying thermal noise in the components. The environment noise uncertainty may be caused by transmissions of other users, either unintentionally or intentionally. Because of noise uncertainty, in practice, it is very difficult to obtain the accurate noise power.

There have been several sensing methods, including the likelihood ratio test (LRT) [13], energy detection method [14, 10, 13, 11, 12, 15], matched filtering (MF)-based method [11, 16, 15, 13] and cyclostationary detection method [17, 18, 19], each of which has different requirements and advantages/disadvantages. Although LRT is proved to be optimal, it is very difficult to use it in practice, because it requires exact channel information, and distributions of source signal and noise. In order to use LRT for detection, we need to obtain the channels, and signal and noise distributions first, which are practically intractable. MF-based method requires perfect knowledge of the channel responses from the primary user to the receiver and accurate synchronization (otherwise its performance will be reduced dramatically) [15, 16]. As mentioned earlier, this may not be possible if the primary users do not cooperate with the secondary users. Cyclostationary detection method needs to know the cyclic frequencies of the primary users, which may not be realistic for many of the spectrum reuse applications. Furthermore, this method demands excessive analog to digital converter (ADC) requirement and signal processing capabilities [11]. Energy detection, unlike the other two methods, does not need any information of the signal to be detected and is robust to unknown dispersed channel and fading. However, energy detection requires perfect knowledge of noise power. Wrong estimation of the noise power leads to SNR wall and high probability of false alarm [10, 11, 12, 15, 20]. As pointed out earlier, the estimated noise power could be quite inaccurate due to noise uncertainty. Thus, the main drawback for the energy detection is its sensitiveness to noise uncertainty [10, 11, 12, 15]. Furthermore, while energy detection is optimal for detecting independent and identically distributed (iid) signal [13], it is not optimal for detecting correlated signal, which is the case for most practical applications.

In this paper, to overcome the shortcoming of energy detection, we propose new methods based on statistical covariances or auto-correlations of the received signal. The statistical covariance matrices or auto-correlations of signal and noise are generally different. Thus this difference is used in the proposed methods to differentiate the signal component from background noise. In practice, there are only limited number of signal samples. Hence, the detection methods are based on the sample covariance matrix. The steps of the proposed methods are as follows. First, the sample covariance matrix of the received signal is computed based on received signal samples. Then two test statistics are extracted from the sample covariance matrix. Finally, a decision on the presence of the signal is made by comparing the ratio of two test statistics with a threshold. Theoretical analysis for the proposed algorithms is given. Detection probability and associated threshold for decision are found based on statistical theory. The methods do not need any information of the signal, the channel and noise power *a priori*. Also, no synchronization is needed. Simulations based on narrowband signals, captured digital television (DTV) signals and multiple antenna signals are presented to evaluate the performance of the proposed methods.

The rest of the paper is organized as follows. The detection algorithms and theoretical analysis are presented in Section II. Section III gives the performance analysis and finds thresholds for the algorithms. Theoretical comparison with the energy detection is also discussed in this section. Simulation results for various types of signals are given in Section IV. Conclusions are drawn in Section V. Finally, a pre-whitening technique is given in the Appendix.

Some notations are used as follows: boldface letters are used to denote matrices and vectors, superscript $(\cdot)^T$ stands for transpose, $\mathbf{I}_q$ denotes the identity matrix of order $q$, $\mathrm{E}[\cdot]$ stands for expectation operation.

## 2 Covariance Based Detections

Let $x_c(t) = s_c(t) + \eta_c(t)$ be the continuous-time received signal, where $s_c(t)$ is the possible primary user's signal and $\eta_c(t)$ is the noise. $\eta_c(t)$ is assumed to be a stationary process satisfying $\mathrm{E}(\eta_c(t)) = 0$, $\mathrm{E}(\eta_c^2(t)) = \sigma_\eta^2$ and $\mathrm{E}(\eta_c(t)\eta_c(t+\tau)) = 0$ for any $\tau \neq 0$. Assume that we are interested in the frequency band with central frequency $f_c$ and bandwidth $W$. We sample the received signal at a sampling rate $f_s$, where $f_s \geq W$. Let $T_s = 1/f_s$ be the sampling period. For notation simplicity, we define $x(n) \triangleq x_c(nT_s)$, $s(n) \triangleq s_c(nT_s)$ and $\eta(n) \triangleq \eta_c(nT_s)$. There are two hypothesizes: $\mathcal{H}_0$, the signal does not exist; and $\mathcal{H}_1$, the signal exists. The received signal samples under the two hypothesizes are given respectively as follows [11, 16, 12, 15]:

$$\mathcal{H}_0: \quad x(n) = \eta(n), \quad (1)$$
$$\mathcal{H}_1: \quad x(n) = s(n) + \eta(n), \quad (2)$$

where $s(n)$ is the transmitted signal samples passed through a wireless channel consisting of path loss, multipath fading and time dispersion effects, and $\eta(n)$ is the white noise which is independent and identically distributed (iid), and with mean zero and variance $\sigma_\eta^2$. Note that $s(n)$ can be the superposition of the received signals from multiple primary users. No synchronization is needed here.

Two probabilities are of interest for spectrum sensing: probability of detection, $P_d$, which defines, at the hypothesis $\mathcal{H}_1$, the probability of the sensing algorithm having detected the presence of the primary signal; and probability of false alarm, $P_{fa}$, which defines, at the hypothesis $\mathcal{H}_0$, the probability of the sensing algorithm claiming the presence of the primary signal.

### 2.1 Covariance Absolute Value (CAV) Detection

Let us consider $L$ consecutive samples and define the following vectors:

$$\mathbf{x}(n) = \begin{bmatrix} x(n) & x(n-1) & \cdots & x(n-L+1) \end{bmatrix}^T \quad (3)$$
$$\mathbf{s}(n) = \begin{bmatrix} s(n) & s(n-1) & \cdots & s(n-L+1) \end{bmatrix}^T, \quad (4)$$
$$\boldsymbol{\eta}(n) = \begin{bmatrix} \eta(n) & \eta(n-1) & \cdots & \eta(n-L+1) \end{bmatrix}^T. \quad (5)$$

The parameter $L$ is called smoothing factor in the following. Considering the statistical covariance matrices of the signal and noise defined as

$$\mathbf{R}_x = \mathrm{E}[\mathbf{x}(n)\mathbf{x}^T(n)], \quad (6)$$
$$\mathbf{R}_s = \mathrm{E}[\mathbf{s}(n)\mathbf{s}^T(n)], \quad (7)$$

we can verify that

$$\mathbf{R}_x = \mathbf{R}_s + \sigma_\eta^2 \mathbf{I}_L. \quad (8)$$

If the signal $s(n)$ is not present, $\mathbf{R}_s = 0$. Hence the off-diagonal elements of $\mathbf{R}_x$ are all zeros. If there is signal and the signal samples are correlated, $\mathbf{R}_s$ is not a diagonal matrix. Hence, some of the off-diagonal elements of $\mathbf{R}_x$ should be non-zeros. Denote $r_{nm}$ as the element of matrix $\mathbf{R}_x$ at the $n$th row and $m$th column, and let

$$T_1 = \frac{1}{L} \sum_{n=1}^{L} \sum_{m=1}^{L} |r_{nm}|, \quad (9)$$

$$T_2 = \frac{1}{L} \sum_{n=1}^{L} |r_{nn}|. \quad (10)$$

Then, if there is no signal, $T_1/T_2 = 1$. If the signal is present, $T_1/T_2 > 1$. Hence, the ratio $T_1/T_2$ can be used to detect the presence of the signal.

In practice, the statistical covariance matrix can only be calculated using a limited number of signal samples. Define the sample auto-correlations of the received signal as

$$\lambda(l) = \frac{1}{N_s} \sum_{m=0}^{N_s-1} x(m)x(m-l), \ l = 0, 1, \cdots, L-1, \quad (11)$$

where $N_s$ is the number of available samples. The statistical covariance matrix $\mathbf{R}_x$ can be approximated by the sample covariance matrix defined as

$$\hat{\mathbf{R}}_x(N_s) = \begin{bmatrix} \lambda(0) & \lambda(1) & \cdots & \lambda(L-1) \\ \lambda(1) & \lambda(0) & \cdots & \lambda(L-2) \\ \vdots & \vdots & \vdots & \vdots \\ \lambda(L-1) & \lambda(L-2) & \cdots & \lambda(0) \end{bmatrix}. \quad (12)$$

Note that the sample covariance matrix is symmetric and Toeplitz. Based on the sample covariance matrix, we propose the following signal detection method.

**Algorithm 1** *The covariance absolute value (CAV) detection algorithm*

*Step 1. Sample the received signal as described above.*

*Step 2. Choose a smoothing factor $L$ and a threshold $\gamma_1$, where $\gamma_1$ should be chosen to meet the requirement for the probability of false alarm. This will be discussed in the next section.*

*Step 3. Compute the auto-correlations of the received signal $\lambda(l)$, $l = 0, 1, \cdots, L-1$, and form the sample covariance matrix.*

*Step 4. Compute*

$$T_1(N_s) = \frac{1}{L} \sum_{n=1}^{L} \sum_{m=1}^{L} |r_{nm}(N_s)|, \quad (13)$$

$$T_2(N_s) = \frac{1}{L} \sum_{n=1}^{L} |r_{nn}(N_s)|, \quad (14)$$

where $r_{nm}(N_s)$ are the elements of the sample covariance matrix $\hat{\mathbf{R}}_x(N_s)$.

*Step 5. Determine the presence of the signal based on $T_1(N_s)$, $T_2(N_s)$ and the threshold $\gamma_1$, i.e., if $T_1(N_s)/T_2(N_s) > \gamma_1$, signal exists; otherwise, signal does not exist.*

**Remark**. The statistics in the algorithm can be calculated directly from the auto-correlations $\lambda(l)$. However, for better understanding and easing the mathematical derivation for the pre-whitening later in Appendix A, here we choose to use the covariance matrix expression.

## 2.2 Theoretical Analysis for CAV Algorithm

The proposed method only uses the received signal samples. It does not need any information of the signal, the channel and noise power as *a priori*. Also, no synchronization is needed.

The validity of the proposed CAV algorithm relies on the assumption that the signal samples are correlated, that is, $\mathbf{R}_s$ is not a diagonal matrix (some of the off-diagonal elements of $\mathbf{R}_s$ should be non-zeros). Obviously, if the signal samples $s(n)$ are iid, then $\mathbf{R}_s = \sigma_s^2 \mathbf{I}_L$. In this case, the assumption is invalid and the algorithm cannot detect the presence of the signal.

However, usually the signal samples should be correlated due to the following reasons.

(1) The signal is oversampled. Let $T_0$ be the Nyquist sampling period of the signal $s_c(t)$ and $s_c(nT_0)$ be the sampled signal based on the Nyquist sampling rate. Based on the sampling theorem, the signal $s_c(t)$ can be expressed as

$$s_c(t) = \sum_{n=-\infty}^{\infty} s_c(nT_0)g(t - nT_0), \quad (15)$$

where $g(t)$ is an interpolation function. Hence, the signal samples $s(n) = s_c(nT_s)$ are only related to $s_c(nT_0)$. If the sampling rate at the receiver $f_s > 1/T_0$, that is, $T_s < T_0$, then $s(n) = s_c(nT_s)$ must be correlated. An example of this is the narrowband signal such as the wireless microphone signal. In a 6 MHz bandwidth TV band, a wireless microphone signal only occupies about 200 KHz. When we sample the received signal with sampling rate not lower than 6 MHz, the wireless microphone signal is actually over-sampled and therefore highly correlated.

(2) The propagation channel has time dispersion, thus the actually signal component at the receiver is given by

$$s_c(t) = \int_{-\infty}^{\infty} h(\tau)s_0(t - \tau) d\tau, \quad (16)$$

where $s_0(t)$ is the original transmitted signal and $h(t)$ is the response of the time dispersive channel. Since the sampling period $T_s$ is usually very small, the integration (16) can be approximated as

$$s_c(t) \approx T_s \sum_{k=-\infty}^{\infty} h(kT_s)s_0(t-kT_s). \quad (17)$$

Hence,

$$s_c(nT_s) \approx T_s \sum_{k=K_0}^{K_1} h(kT_s)s_0((n-k)T_s), \quad (18)$$

where $[K_0 T_s, K_1 T_s]$ is the support of the channel response $h(t)$, that is, $h(t) = 0$ for $t \notin [K_0 T_s, K_1 T_s]$. For time dispersive channel, $K_1 > K_0$, thus the signal samples $s(nT_s)$ are correlated even if the original signal samples $s_0(nT_s)$ could be iid.

(3) The original signal is correlated. In this case, even if the channel is a flat fading channel and no oversampling, the received signal samples are correlated.

Another assumption for the algorithm is that the noise samples are iid. This is usually true if no filtering is used. However, if a narrowband filter is used at the receiver, sometimes the noise samples will be correlated. To deal with this case, we need to pre-whiten the noise samples or pre-transform the covariance matrix. A method is given in Appendix A to solve this problem.

The computational complexity of the algorithm is as follows. Computing the auto-correlations of the received signal requires about $LN_s$ multiplications and additions. Computing $T_1(N_s)$ and $T_2(N_s)$ requires about $L^2$ additions. Therefore, the total number of multiplications and additions is about $LN_s + L^2$.

### 2.3 Generalized Covariance Based Algorithms

Based on the same principle as CAV, generalized covariance based methods can be designed to detect the signal. Let $\psi_1$ and $\psi_2$ be two non-negative functions with multiple variables. Assume that

$$\psi_1(\mathbf{a}) > 0, \text{ for } \mathbf{a} \neq \mathbf{0}; \ \psi_1(\mathbf{0}) = 0;$$

$$\psi_2(\mathbf{b}) > 0, \text{ for } \mathbf{b} \neq \mathbf{0}; \ \psi_2(\mathbf{0}) = 0.$$

Then, the following method can be used for signal detection.

**Algorithm 2** *Generalized Covariance based Detection*

*Step 1. Sample the received signal as described above.*

*Step 2. Choose a smoothing factor $L$ and a threshold $\gamma_2$, where $\gamma_2$ should be chosen to meet the requirement for the probability of false alarm.*

*Step 3. Compute the sample covariance matrix $\hat{\mathbf{R}}_x(N_s)$.*

*Step 4. Compute*

$$T_4(N_s) = \psi_2(r_{nn}(N_s), \ n = 1, \cdots, L), \quad (19)$$

$$T_3(N_s) = T_4(N_s) + \psi_1(r_{nm}(N_s), \ n \neq m). \quad (20)$$

*Step 5. Determine the presence of the signal based on $T_3(N_s)$, $T_4(N_s)$ and the threshold $\gamma_2$. That is, if $T_3(N_s)/T_4(N_s) > \gamma_2$, signal exists; otherwise, signal does not exist.*

Obviously the CAV algorithm is a special case of the generalized method when $\psi_1$ and $\psi_2$ are absolute summation functions. As another example, we can choose $\psi_1(\mathbf{a}) = \mathbf{a}^T \mathbf{a}$ and $\psi_2(\mathbf{b}) = \mathbf{b}^T \mathbf{b}$. For this choice,

$$T_3(N_s) = \frac{1}{L} \sum_{n=1}^{L} \sum_{m=1}^{L} |r_{nm}(N_s)|^2, \quad (21)$$

$$T_4(N_s) = \frac{1}{L} \sum_{n=1}^{L} |r_{nn}(N_s)|^2. \quad (22)$$

### 2.4 Spectrum sensing Using Multiple Antennas

Multiple antenna systems have been widely used to increase the channel capacity or improve the transmission reliability in wireless communications. In the following, we assume that there are $M > 1$ antennas at the receiver, and exploit the received signals from these antennas for spectrum sensing. In this case, the received signal at antenna $i$ is given by

$$\mathcal{H}_0: \ x_i(n) = \eta_i(n), \quad (23)$$
$$\mathcal{H}_1: \ x_i(n) = s_i(n) + \eta_i(n). \quad (24)$$

In hypothesis $\mathcal{H}_1$, $s_i(n)$ is the signal component received by antenna $i$. Since all $s_i(n)$ are generated from the same source signal, thus $s_i(n)$ are correlated for $i$. It is assumed that $\eta_i(n)$'s are iid for $n$ and $i$.

Let us combine all the signals from the $M$ antennas and define the following vectors:

$$\mathbf{x}(n) = [\ x_1(n) \ \cdots \ x_M(n) \ x_1(n-1) \ \cdots \ x_M(n-1)$$
$$\cdots \ x_1(n-L+1) \cdots \ x_M(n-L+1)\ ]^T, \quad (25)$$
$$\mathbf{s}(n) = [\ s_1(n) \ \cdots \ s_M(n) \ s_1(n-1) \ \cdots \ s_M(n-1)$$
$$\cdots \ s_1(n-L+1) \cdots \ s_M(n-L+1)\ ]^T, \quad (26)$$
$$\boldsymbol{\eta}(n) = [\ \eta_1(n) \ \cdots \ \eta_M(n) \ \eta_1(n-1) \ \cdots \ \eta_M(n-1)$$
$$\cdots \ \eta_1(n-L+1) \cdots \ \eta_M(n-L+1)\ ]^T. \quad (27)$$

Note that equations (3) to (5) are a special case ($M = 1$) of the above equations. Defining the statistical covariance matrices using the same way as those in (6) and (7), we obtain

$$\mathbf{R}_x = \mathbf{R}_s + \sigma_\eta^2 \mathbf{I}_{ML}. \quad (28)$$

Except the different matrix dimensions, the equation above is the same as equation (8). Hence, the CAV algorithm and generalized covariance based method described above can be directly used for multiple antenna case.

Let $s_0(n)$ be the source signal. The received signal at antenna $i$ is

$$s_i(n) = \sum_{k=0}^{N_i} h_i(k)s_0(n-k) + \eta_i(n), \ i = 1, 2, \cdots, M, \quad (29)$$

where $h_i(k)$ is the channel responses from the source user to antenna $i$ at the receiver. Define

$$\mathbf{h}(n) = [h_1(n), h_2(n), \cdots, h_M(n)]^T, \quad (30)$$

$$\mathbb{H} = \begin{bmatrix} \mathbf{h}(0) & \cdots & \cdots & \mathbf{h}(N) & \cdots & 0 \\ & \ddots & & & \ddots & \\ 0 & \cdots & \mathbf{h}(0) & \cdots & \cdots & \mathbf{h}(N) \end{bmatrix}, \quad (31)$$

where $N = \max_i(N_i)$ and $h_i(n)$ is zero-padded if $N_i < N$. Note that the dimension of $\mathbb{H}$ is $ML \times (N+L)$. We have

$$\mathbf{R}_s = \mathbb{H} \mathbf{R}_{s_0} \mathbb{H}^T, \quad (32)$$

where $\mathbf{R}_{s_0} = \mathrm{E}(\hat{\mathbf{s}}_0 \hat{\mathbf{s}}_0^T)$ is the statistical covariance matrix of the source signal, where

$$\hat{\mathbf{s}}_0 = \begin{bmatrix} s_0(n) & s_0(n-1) & \cdots & s_0(n-N-L+1) \end{bmatrix}^T. \quad (33)$$

Note that the received signals at different antennas are correlated. Hence, using multiple antennas increase the correlations among the signal samples at the receiver and make the algorithms valid at all cases. In fact, at the worst case when all the channels are flat-fading, that is, $N_1 = N_2 = \cdots = N_M = 0$, and the source signal sample $s_0(n)$ is iid, we have $\mathbf{R}_s = \sigma_s^2 \mathbb{H} \mathbb{H}^T$, where $\mathbb{H}$ is a $ML \times L$ matrix defined above. Obviously, $\mathbf{R}_s$ is not a diagonal matrix and the algorithms can work.

## 3 Performance Analysis and Threshold Determination

For a good detection algorithm, $P_d$ should be high and $P_{fa}$ should be low. The choice of the threshold $\gamma$ is a compromise between the $P_d$ and $P_{fa}$. Since we have no information on the signal (actually we even do not know if there is signal or not), it is difficult to set the threshold based on the $P_d$. Hence, usually we choose the threshold based on the $P_{fa}$. The steps are as follows. First we set a value for $P_{fa}$. Then we find a threshold $\gamma$ to meet the required $P_{fa}$. To find the threshold based on the required $P_{fa}$, we can use either theoretical derivation or computer simulation. If simulation is used to find the threshold, we can generate white Gaussian noises as the input (no signal) and adjust the threshold to meet the $P_{fa}$ requirement. Note that the threshold here is related to the number of samples used for computing the sample auto-correlations and the smoothing factor $L$, but not related to the noise power. If theoretical derivation is used, we need to find the statistical distribution of $T_1(N_s)/T_2(N_s)$, which is generally a difficult task. In this section, using central limit theorem, we will find the approximations for the distribution of this random variable and provide the theoretical estimations for the two probabilities, $P_d$, $P_{fa}$, as well as the threshold associated with these probabilities.

### 3.1 Statistics Computation

Based on the symmetric property of the covariance matrix, we can rewrite $T_1(N_s)$ and $T_2(N_s)$ in (13) and (14) as

$$T_1(N_s) = \lambda(0) + \frac{2}{L} \sum_{l=1}^{L-1} (L-l)|\lambda(l)|, \quad (34)$$

$$T_2(N_s) = \lambda(0). \quad (35)$$

Define

$$\mathbf{X}_l = \begin{bmatrix} x(N_s - 1 - l) & \cdots & x(-l) \end{bmatrix}^T, \quad (36)$$

$$\mathbf{S}_l = \begin{bmatrix} s(N_s - 1 - l) & \cdots & s(-l) \end{bmatrix}^T, \quad (37)$$

$$\boldsymbol{\eta}_l = \begin{bmatrix} \eta(N_s - 1 - l) & \cdots & \eta(-l) \end{bmatrix}^T. \quad (38)$$

Let the normalized correlation among the signal samples be

$$\alpha_l = \mathrm{E}[s(n)s(n-l)]/\sigma_s^2, \quad (39)$$

where $\sigma_s^2$ is the signal power, $\sigma_s^2 = \mathrm{E}[s^2(n)]$. $|\alpha_l|$ defines the correlation strength among the signal samples, here $0 \leqslant |\alpha_l| \leqslant 1$. Based on the notations, we have

$$\lambda(l) = \frac{1}{N_s} \mathbf{X}_0^T \mathbf{X}_l = \frac{1}{N_s} (\mathbf{S}_0^T + \boldsymbol{\eta}_0^T)(\mathbf{S}_l + \boldsymbol{\eta}_l)$$

$$= \frac{1}{N_s} (\mathbf{S}_0^T \mathbf{S}_l + \mathbf{S}_0^T \boldsymbol{\eta}_l + \boldsymbol{\eta}_0^T \mathbf{S}_l + \boldsymbol{\eta}_0^T \boldsymbol{\eta}_l). \quad (40)$$

Obviously,

$$\mathrm{E}(\lambda(0)) = \sigma_s^2 + \sigma_\eta^2, \quad (41)$$

$$\mathrm{E}(\lambda(l)) = \alpha_l \sigma_s^2, \ l = 1, 2, \cdots, L-1. \quad (42)$$

Now we need to find the variance of $\lambda(l)$. Since

$$\lambda^2(l) = \frac{1}{N_s^2} (\mathbf{S}_0^T \mathbf{S}_l + \mathbf{S}_0^T \boldsymbol{\eta}_l + \boldsymbol{\eta}_0^T \mathbf{S}_l + \boldsymbol{\eta}_0^T \boldsymbol{\eta}_l)^2$$

$$= \frac{1}{N_s^2} \Big[ (\mathbf{S}_0^T \mathbf{S}_l)^2 + (\mathbf{S}_0^T \boldsymbol{\eta}_l)^2 + (\boldsymbol{\eta}_0^T \mathbf{S}_l)^2 + (\boldsymbol{\eta}_0^T \boldsymbol{\eta}_l)^2$$
$$+ 2(\mathbf{S}_0^T \mathbf{S}_l)(\mathbf{S}_0^T \boldsymbol{\eta}_l) + 2(\mathbf{S}_0^T \mathbf{S}_l)(\boldsymbol{\eta}_0^T \mathbf{S}_l)$$
$$+ 2(\mathbf{S}_0^T \mathbf{S}_l)(\boldsymbol{\eta}_0^T \boldsymbol{\eta}_l) + 2(\mathbf{S}_0^T \boldsymbol{\eta}_l)(\boldsymbol{\eta}_0^T \mathbf{S}_l)$$
$$+ 2(\mathbf{S}_0^T \boldsymbol{\eta}_l)(\boldsymbol{\eta}_0^T \boldsymbol{\eta}_l) + 2(\boldsymbol{\eta}_0^T \mathbf{S}_l)(\boldsymbol{\eta}_0^T \boldsymbol{\eta}_l) \Big], \quad (43)$$

it can be verified that

$$\mathrm{E}\left((\mathbf{S}_0^T \boldsymbol{\eta}_l)^2\right) = \mathrm{E}\left((\boldsymbol{\eta}_0^T \mathbf{S}_l)^2\right) = N_s \sigma_s^2 \sigma_\eta^2, \quad (44)$$

$$\mathrm{E}\left((\boldsymbol{\eta}_0^T \boldsymbol{\eta}_0)^2\right) = (N_s^2 + 2N_s)\sigma_\eta^4, \quad (45)$$

$$\mathrm{E}\left((\boldsymbol{\eta}_0^T \boldsymbol{\eta}_l)^2\right) = N_s \sigma_\eta^4, \ l = 1, \cdots, L-1, \quad (46)$$

$$\mathrm{E}\left((\mathbf{S}_0^T \mathbf{S}_l)(\mathbf{S}_0^T \boldsymbol{\eta}_l)\right) = \mathrm{E}\left((\mathbf{S}_0^T \mathbf{S}_l)(\boldsymbol{\eta}_0^T \mathbf{S}_l)\right)$$
$$= \mathrm{E}\left((\mathbf{S}_0^T \boldsymbol{\eta}_l)(\boldsymbol{\eta}_0^T \boldsymbol{\eta}_l)\right)$$
$$= \mathrm{E}\left((\boldsymbol{\eta}_0^T \mathbf{S}_l)(\boldsymbol{\eta}_0^T \boldsymbol{\eta}_l)\right) = 0, \quad (47)$$

$$\mathrm{E}\left((\mathbf{S}_0^T \mathbf{S}_0)(\boldsymbol{\eta}_0^T \boldsymbol{\eta}_0)\right) = N_s^2 \sigma_s^2 \sigma_\eta^2, \quad (48)$$

$$\mathrm{E}\left((\mathbf{S}_0^T \mathbf{S}_l)(\boldsymbol{\eta}_0^T \boldsymbol{\eta}_l)\right) = 0, \ l = 1, 2, \cdots, L-1, \quad (49)$$

$$\mathrm{E}\left((\mathbf{S}_0^T \boldsymbol{\eta}_l)(\boldsymbol{\eta}_0^T \mathbf{S}_l)\right) = \alpha_{2l}(N_s - l)\sigma_s^2 \sigma_\eta^2, \quad (50)$$

$$\mathrm{E}\left(\mathbf{S}_0^T \mathbf{S}_l\right) = \alpha_l \sigma_s^2. \quad (51)$$

Based on these results, we can easily obtain the following two lemmas.

**Lemma 1** *When there is no signal, we have*

$$\mathrm{E}(\lambda(0)) = \sigma_\eta^2, \ \mathrm{Var}(\lambda(0)) = \frac{2}{N_s}\sigma_\eta^4, \quad (52)$$

$$\mathrm{E}(\lambda(l)) = 0, \ \mathrm{Var}(\lambda(l)) = \frac{1}{N_s}\sigma_\eta^4, \ l = 1,\cdots,L-1. \quad (53)$$

**Lemma 2** *When there is signal, we have*

$$\begin{aligned}
\mathrm{E}(\lambda(0)) &= \sigma_s^2 + \sigma_\eta^2, & (54)\\
\mathrm{Var}(\lambda(0)) &= \mathrm{Var}\left(\frac{1}{N_s}\mathbf{S}_0^T\mathbf{S}_0\right) + \frac{2\sigma_\eta^2}{N_s}\left(2\sigma_s^2 + \sigma_\eta^2\right), & (55)\\
\mathrm{E}(\lambda(l)) &= \alpha_l \sigma_s^2, & (56)\\
\mathrm{Var}(\lambda(l)) &= \mathrm{Var}\left(\frac{1}{N_s}\mathbf{S}_0^T\mathbf{S}_l\right) \\
&\quad + \frac{\sigma_\eta^2}{N_s}\left(\sigma_\eta^2 + 2\sigma_s^2 + \frac{2(N_s - l)\alpha_{2l}}{N_s}\sigma_s^2\right) & (57)\\
& l = 1,\cdots,L-1
\end{aligned}$$

Note that $\mathrm{Var}\left(\frac{1}{N_s}\mathbf{S}_0^T\mathbf{S}_l\right)$, $l = 0,1,\cdots,L-1$, depends on signal properties.

For simplicity, we denote $\mathrm{E}(\lambda(l))$ by $\Theta_l$ and $\mathrm{Var}(\lambda(l))$ by $\Delta_l$. Note that usually $N_s$ is very large. Based on central limit theorem, $\lambda(l)$ can be approximated by the Gaussian distribution.

**Lemma 3** *When the signal is not present, we have*

$$\mathrm{E}(|\lambda(l)|) = \sqrt{\frac{2}{\pi N_s}}\sigma_\eta^2, \ l = 1,2,\cdots,L-1. \quad (58)$$

*When the signal is present, we have*

$$\begin{aligned}
\mathrm{E}(|\lambda(l)|) &= \sqrt{\frac{2\Delta_l}{\pi}}\left(2 - e^{-\frac{\Theta_l^2}{2\Delta_l}}\right)\\
&\quad + |\Theta_l|\left(1 - \sqrt{\frac{2}{\pi}}\int_{|\Theta_l|/\sqrt{\Delta_l}}^{+\infty} e^{-\frac{u^2}{2}}\,du\right) & (59)\\
& l = 1,2,\cdots,L-1.
\end{aligned}$$

*For large $N_s$ and low SNR,*

$$\begin{aligned}
\mathrm{E}(|\lambda(l)|) &\approx \sqrt{\frac{2}{\pi N_s}}(\sigma_s^2 + \sigma_\eta^2)\left(2 - e^{-\frac{\tau_l^2}{2}}\right)\\
&\quad + |\Theta_l|\left(1 - \sqrt{\frac{2}{\pi}}\int_{\tau_l}^{+\infty} e^{-\frac{u^2}{2}}\,du\right), & (60)\\
& l = 1,2,\cdots,L-1,
\end{aligned}$$

*where*

$$\tau_l = \frac{|\alpha_l|\mathrm{SNR}\sqrt{N_s}}{1 + \mathrm{SNR}}, \ \mathrm{SNR} = \frac{\sigma_s^2}{\sigma_\eta^2}. \quad (61)$$

**Proof.** Based on the central limit theorem, we have

$$\begin{aligned}
\mathrm{E}(|\lambda(l)|) &= \frac{1}{\sqrt{2\pi\Delta_l}}\int_{-\infty}^{+\infty} |u| e^{-\frac{(u-\Theta_l)^2}{2\Delta_l}}\,du\\
&= \frac{1}{\sqrt{2\pi}}\int_{-\infty}^{+\infty} |\sqrt{\Delta_l}u + \Theta_l| e^{-\frac{u^2}{2}}\,du\\
&= \frac{1}{\sqrt{2\pi}}\int_{-\infty}^{-\Theta_l/\sqrt{\Delta_l}} (-\sqrt{\Delta_l}u - \Theta_l) e^{-\frac{u^2}{2}}\,du\\
&\quad + \frac{1}{\sqrt{2\pi}}\int_{-\Theta_l/\sqrt{\Delta_l}}^{+\infty} (\sqrt{\Delta_l}u + \Theta_l) e^{-\frac{u^2}{2}}\,du\\
&= \sqrt{\frac{2\Delta_l}{\pi}}\int_0^{+\infty} u e^{-\frac{u^2}{2}}\,du + \sqrt{\frac{2\Delta_l}{\pi}}\int_{-\Theta_l/\sqrt{\Delta_l}}^0 u e^{-\frac{u^2}{2}}\,du\\
&\quad + \frac{2\Theta_l}{\sqrt{2\pi}}\int_{-\Theta_l/\sqrt{\Delta_l}}^0 e^{-\frac{u^2}{2}}\,du\\
&= \sqrt{\frac{2\Delta_l}{\pi}}\left(2 - e^{-\frac{\Theta_l^2}{2\Delta_l}}\right) + \frac{2\Theta_l}{\sqrt{2\pi}}\int_{-\Theta_l/\sqrt{\Delta_l}}^0 e^{-\frac{u^2}{2}}\,du\\
&= \sqrt{\frac{2\Delta_l}{\pi}}\left(2 - e^{-\frac{\Theta_l^2}{2\Delta_l}}\right)\\
&\quad + |\Theta_l|\left(1 - \sqrt{\frac{2}{\pi}}\int_{|\Theta_l|/\sqrt{\Delta_l}}^{+\infty} e^{-\frac{u^2}{2}}\,du\right). \quad (62)
\end{aligned}$$

For large $N_s$ and low SNR,

$$\Delta_l \approx \frac{(\sigma_s^2 + \sigma_\eta^2)^2}{N_s}, \ \frac{|\Theta_l|}{\sqrt{\Delta_l}} \approx \tau_l.$$

So, we obtain (60).

When there is no signal, $\Theta_l = 0$ and $\Delta_l = \frac{1}{N_s}\sigma_\eta^4$. Hence, we obtain equation (58).

$\square$

**Theorem 1** *When there is no signal, we have*

$$\begin{aligned}
\mathrm{E}(T_1(N_s)) &= \left(1 + (L-1)\sqrt{\frac{2}{\pi N_s}}\right)\sigma_\eta^2, & (63)\\
\mathrm{E}(T_2(N_s)) &= \sigma_\eta^2, & (64)\\
\mathrm{Var}(T_2(N_s)) &= \frac{2}{N_s}\sigma_\eta^4. & (65)
\end{aligned}$$

*When there is signal, and for large $N_s$, we have*

$$\mathrm{E}(T_1(N_s)) \approx \sigma_s^2$$
$$+ \sigma_\eta^2 + \frac{2\sigma_s^2}{L}\sum_{l=1}^{L-1}(L-l)|\alpha_l|\left(1 - \sqrt{\frac{2}{\pi}}\int_{\tau_l}^{+\infty} e^{-\frac{u^2}{2}}\,du\right)$$
$$+ \frac{2(\sigma_s^2 + \sigma_\eta^2)}{L}\sum_{l=1}^{L-1}(L-l)\sqrt{\frac{2}{\pi N_s}}\left(2 - e^{-\frac{\tau_l^2}{2}}\right), \quad (66)$$

$$\mathrm{E}(T_2(N_s)) = \sigma_s^2 + \sigma_\eta^2, \quad (67)$$

$$\mathrm{Var}(T_2(N_s)) = \mathrm{Var}\left(\frac{1}{N_s}\mathbf{S}_0^T\mathbf{S}_0\right) + \frac{2\sigma_\eta^2}{N_s}\left(2\sigma_s^2 + \sigma_\eta^2\right). \quad (68)$$

*For large $N_s$, $T_1(N_s)$ and $T_2(N_s)$ approach to Gaussian distributions.*

**Proof.** Equations (63) to 68) are direct results from Lemma 1 to 3. Noting that $\lambda(l)$ is a summation of $N_s$ random variables, when $N_s$ is large, based on the central limit theorem it can be approximated by Gaussian distributions. From the definition of $T_1(N_s)$ and $T_2(N_s)$, we know that they also approach to Gaussian distributions.

$\square$

## 3.2 Detection Probability and the Associated Threshold

From the theorem above, we have

$$\lim_{N_s \to \infty} \mathrm{E}(T_1(N_s)) = \sigma_s^2 + \sigma_\eta^2 + \frac{2\sigma_s^2}{L}\sum_{l=1}^{L-1}(L-l)|\alpha_l|. \quad (69)$$

For simplicity, we denote

$$\Upsilon_L \triangleq \frac{2}{L}\sum_{l=1}^{L-1}(L-l)|\alpha_l|, \quad (70)$$

which is the overall correlation strength among the consecutive $L$ samples. When there is no signal, we have

$$\begin{aligned} T_1(N_s)/T_2(N_s) &\approx \mathrm{E}(T_1(N_s))/\mathrm{E}(T_2(N_s)) \\ &= 1 + (L-1)\sqrt{\frac{2}{\pi N_s}}. \end{aligned} \quad (71)$$

Note that this ratio approaches to 1 as $N_s$ approaches to infinite. Also note that the ratio is not related to the noise power (variance). On the other hand, when there is signal (signal plus noise case), we have

$$\begin{aligned} T_1(N_s)/T_2(N_s) &\approx \mathrm{E}(T_1(N_s))/\mathrm{E}(T_2(N_s)) \\ &\approx 1 + \frac{\sigma_s^2}{\sigma_s^2 + \sigma_\eta^2}\Upsilon_L \quad (72) \\ &= 1 + \frac{\mathrm{SNR}}{\mathrm{SNR}+1}\Upsilon_L. \quad (73) \end{aligned}$$

Here the ratio approaches to a number larger than 1 as $N_s$ approaches to infinite. The number is determined by the correlation strength among the signal samples and the SNR. Hence, for any fixed SNR, if there are sufficiently large number of samples, we can always differentiate if there is signal or not based on the ratio.

However, in practice we have only limited number of samples. So, we need to evaluate the performance at fixed $N_s$. First we analyze the $P_{fa}$ at hypothesis $\mathcal{H}_0$. The probability of false alarm for the CAV algorithm is

$$\begin{aligned} P_{fa} &= P(T_1(N_s) > \gamma_1 T_2(N_s)) \\ &= P\left(T_2(N_s) < \frac{1}{\gamma_1}T_1(N_s)\right) \\ &\approx P\left(T_2(N_s) < \frac{1}{\gamma_1}\left(1+(L-1)\sqrt{\frac{2}{N_s\pi}}\right)\sigma_\eta^2\right) \\ &= P\left(\frac{T_2(N_s)-\sigma_\eta^2}{\sqrt{\frac{2}{N_s}}\sigma_\eta^2} < \frac{\frac{1}{\gamma_1}\left(1+(L-1)\sqrt{\frac{2}{N_s\pi}}\right)-1}{\sqrt{2/N_s}}\right) \\ &\approx 1 - \mathrm{Q}\left(\frac{\frac{1}{\gamma_1}\left(1+(L-1)\sqrt{\frac{2}{N_s\pi}}\right)-1}{\sqrt{2/N_s}}\right) \end{aligned}$$

where

$$\mathrm{Q}(t) = \frac{1}{\sqrt{2\pi}}\int_t^{+\infty} e^{-u^2/2}\mathrm{d}u. \quad (74)$$

For a given $P_{fa}$, the associated threshold should be chosen such that

$$\frac{\frac{1}{\gamma_1}\left(1+(L-1)\sqrt{\frac{2}{N_s\pi}}\right)-1}{\sqrt{2/N_s}} = -\mathrm{Q}^{-1}(P_{fa}). \quad (75)$$

That is,

$$\gamma_1 = \frac{1+(L-1)\sqrt{\frac{2}{N_s\pi}}}{1-\mathrm{Q}^{-1}(P_{fa})\sqrt{\frac{2}{N_s}}}. \quad (76)$$

Note that here the threshold is not related to noise power and SNR. After the threshold is set, we now calculate the probability of detection at various SNR. For the given threshold $\gamma_1$, when signal presents,

$$\begin{aligned} P_d &= P(T_1(N_s) > \gamma_1 T_2(N_s)) \\ &= P\left(T_2(N_s) < \frac{1}{\gamma_1}T_1(N_s)\right) \\ &\approx P\left(T_2(N_s) < \frac{1}{\gamma_1}\mathrm{E}(T_1(N_s))\right) \\ &= P\left(\frac{T_2(N_s)-\sigma_s^2-\sigma_\eta^2}{\sqrt{\mathrm{Var}(T_2(N_s))}} < \frac{\frac{1}{\gamma_1}\mathrm{E}(T_1(N_s))-\sigma_s^2-\sigma_\eta^2}{\sqrt{\mathrm{Var}(T_2(N_s))}}\right) \\ &= 1 - \mathrm{Q}\left(\frac{\frac{1}{\gamma_1}\mathrm{E}(T_1(N_s))-\sigma_s^2-\sigma_\eta^2}{\sqrt{\mathrm{Var}(T_2(N_s))}}\right). \quad (77) \end{aligned}$$

For very large $N_s$ and low SNR,

$$\mathrm{Var}(T_2(N_s)) \approx \frac{2\sigma_\eta^2}{N_s}\left(2\sigma_s^2+\sigma_\eta^2\right) \approx \frac{2(\sigma_s^2+\sigma_\eta^2)^2}{N_s},$$

$$\mathrm{E}(T_1(N_s)) \approx \sigma_s^2+\sigma_\eta^2+\sigma_s^2\Upsilon_L.$$

Hence, we have a further approximation

$$\begin{aligned} P_d &\approx 1 - \mathrm{Q}\left(\frac{\frac{1}{\gamma_1}+\frac{\Upsilon_L \sigma_s^2}{\gamma_1(\sigma_s^2+\sigma_\eta^2)}-1}{\sqrt{2/N_s}}\right) \\ &= 1 - \mathrm{Q}\left(\frac{\frac{1}{\gamma_1}+\frac{\Upsilon_L \mathrm{SNR}}{\gamma_1(\mathrm{SNR}+1)}-1}{\sqrt{2/N_s}}\right). \quad (78) \end{aligned}$$

Obviously, the $P_d$ increases with the number of samples, $N_s$, the SNR and the correlation strength among the signal samples. Note that $\gamma_1$ is also related to $N_s$ as shown above, and $\lim_{N_s \to \infty} \gamma_1 = 1$. Hence, for fixed SNR, $P_d$ approaches to 1 when $N_s$ approaches to infinite.

For a target pair of $P_d$ and $P_{fa}$, based on (78) and (76), we can find the required number of samples as

$$N_c \approx \frac{2\left(\mathrm{Q}^{-1}(P_{fa}) - \mathrm{Q}^{-1}(P_d) + (L-1)/\sqrt{\pi}\right)^2}{(\Upsilon_L \mathrm{SNR})^2}. \quad (79)$$

For fixed $P_d$ and $P_{fa}$, $N_c$ is only related to the smoothing factor $L$ and the overall correlation strength $\Upsilon_L$. Hence, the best smoothing factor is

$$L_{best} = \min_L \{N_c\}, \quad (80)$$

which is related to the correlation strength among the signal samples.

### 3.3 Comparison with Energy Detection

Energy detection is the basic sensing method, which was first proposed in [14] and further studied by others [10, 11, 12, 15]. It does not need any information of the signal to be detected and is robust to unknown dispersive channel. Energy detection compares the average power of the received signal with the noise power to make a decision. To guarantee a reliable detection, the threshold must be set according to the noise power and the number of samples [11, 10, 12]. On the other hand, the proposed methods do not rely on the noise power to set the threshold (see equation (76)), while keeping other advantages of the energy detection.

Accurate knowledge on the noise power is then the key of the energy detection. Unfortunately, in practice, the noise uncertainty always presents. Due to the noise uncertainty [10, 11, 12], the estimated (or assumed) noise power may be different from the real noise power. Let the estimated noise power be $\hat{\sigma}_\eta^2 = \alpha \sigma_\eta^2$. We define the noise uncertainty factor (in dB) as

$$B = \max\{10 \log_{10} \alpha\}. \quad (81)$$

It is assumed that $\alpha$ (in dB) is evenly distributed in an interval $[-B, B]$ [11]. In practice, the noise uncertainty factor of a receiving device is normally 1 to 2 dB [11, 20]. The environment/interference noise uncertainty can be much higher [11]. When there is noise uncertainty, the energy detection is not effective [10, 11, 12, 20]. Simulation in the next section also shows that the proposed method is much better than the energy detection when noise uncertainty is present. Hence, here we only compare the proposed method with the ideal energy detection (without noise uncertainty).

To compare the performances of the two methods, first we need a criterion. By properly choosing the thresholds, both methods can achieve any given $P_d$ and $P_{fa} > 0$ if sufficiently large number of samples are available. The key point is how many samples are needed to achieve the given $P_d$ and $P_{fa} > 0$. Hence, we choose this as the criterion to compare the two algorithms. For energy detection, the required number of samples is approximately [11]

$$N_e = \frac{2\left(\mathrm{Q}^{-1}(P_{fa}) - \mathrm{Q}^{-1}(P_d)\right)^2}{\mathrm{SNR}^2}. \quad (82)$$

Comparing (79) and (82), if we want $N_c < N_e$, we need

$$\Upsilon_L > 1 + \frac{L-1}{\sqrt{\pi}\left(\mathrm{Q}^{-1}(P_{fa}) - \mathrm{Q}^{-1}(P_d)\right)}. \quad (83)$$

For example, if $P_d = 0.9$ and $P_{fa} = 0.1$, we need $\Upsilon_L > 1 + \frac{L-1}{4.54}$. In conclusion, if the signal samples are highly correlated such that (83) holds, the CAV is better than the ideal energy detection; otherwise, the ideal energy detection is better.

In terms of the computational complexity, the energy detection needs about $N_s$ multiplications and additions. Hence, the computational complexity of the proposed methods is about $L$ times that of the energy detection.

## 4 Simulations and Discussions

In this section, we will give some simulation results for three situations: narrowband signals, captured DTV signals [21] and multiple antenna received signals.

First, we simulate the probabilities of false alarm ($P_{fa}$) because $P_{fa}$ is not related to signal (at $\mathcal{H}_0$, there is no signal at all). We set the target $P_{fa} = 0.1$, and choose $L = 10$ and $N_s = 50000$. We then obtain the thresholds based on the $P_{fa}$, $L$ and $N_s$. The threshold for energy detection is given in [11]. Table 1 gives the simulation results for various cases, where and in the following "EG-x dB" means the energy detection with x-dB noise uncertainty. The $P_{fa}$ for the proposed method and energy detection without noise uncertainty meet the target, but the $P_{fa}$ for the energy detection with noise uncertainty (even as low as 0.5 dB) far exceeds the limit. This means that the energy detection is very unreliable in practical situations with noise uncertainty.

Secondly, we fix the thresholds based on $P_{fa}$ and simulate the probability of detection ($P_d$) for various cases. We consider two signal types as follows.

**(1) Narrowband signals.** FM modulated wireless microphone signal is used here (soft speaker) [22]. The central frequency is $f_c = 100$ MHz. The sampling rate at the receiver is 6 MHz (the same as the TV bandwidth in USA). Figure 1 gives the simulation results (**the corresponding $P_{fa}$ is shown in Table 1**). Note that "CAV-theo" means the theoretical results given in Section III.B. Due to some approximations, the theoretical results does not exactly match the simulated results. The CAV is better than ideal energy detection (without noise uncertainty), which verifies our assertion in Section III. C. The reason is that, as we point out in Section II.B, the source signal is narrowband, and therefore their samples are highly correlated. As shown in the figure, if there is noise uncertainty, the $P_d$ of the energy detection is much worse than that of the proposed method. Figure 2 gives the Receiver Operating Characteristics (ROC) curve ($P_d$ versus $P_{fa}$) at fixed SNR $= -20$dB. The performances of the methods at different sample sizes (sensing times) are given in Figure 3. It is clear that CAV is always better than the ideal energy detection.

| method | EG-2 dB | EG-1.5 dB | EG-1 dB | EG-0.5 dB | EG-0dB | CAV |
|---|---|---|---|---|---|---|
| $P_{fa}$ | 0.497 | 0.496 | 0.490 | 0.470 | 0.102 | 0.099 |

Table 1: Probabilities of false alarm

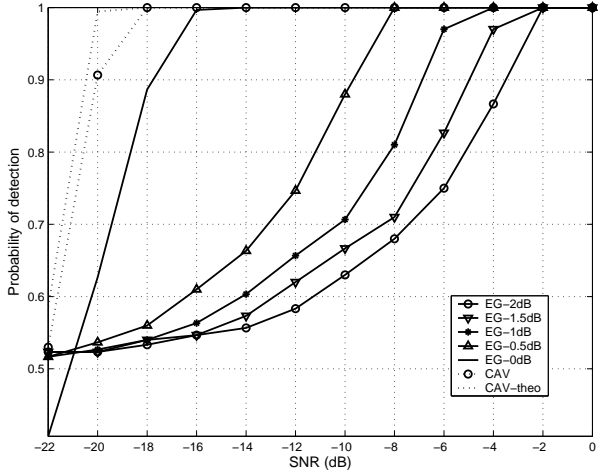

Figure 1: Probability of detection for wireless microphone signal: $N_s = 50000$

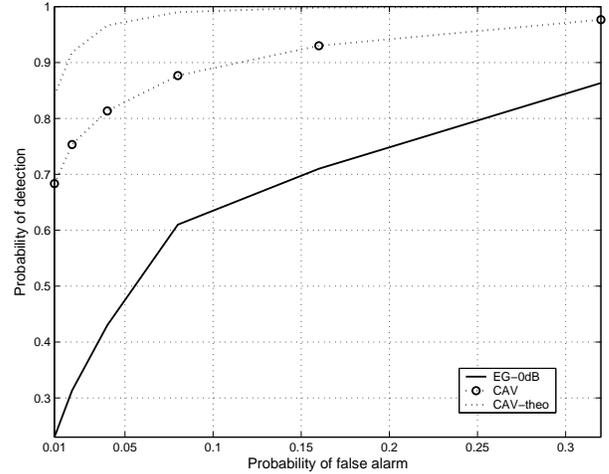

Figure 2: $P_d$ versus $P_{fa}$ for wireless microphone signal: $N_s = 50000$, SNR $= -20$dB

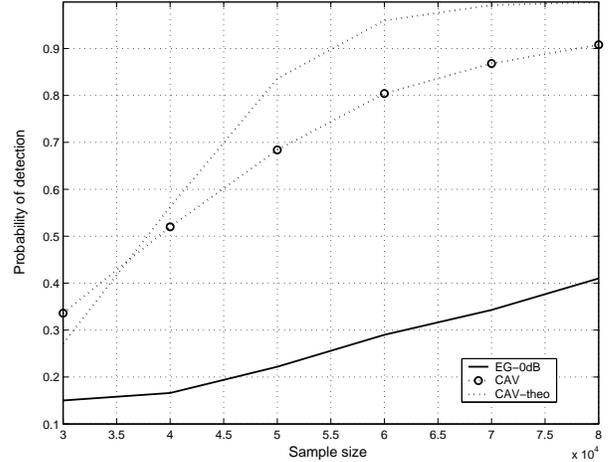

Figure 3: $P_d$ versus sample size ($N_s$) for wireless microphone signal: $P_{fa} = 0.01$, SNR $= -20$dB

To test the impact of the smoothing factor, we fix SNR=-20dB, $P_{fa} = 0.01$, and $N_s = 50000$, and vary the smoothing factor $L$ from 4 to 14. Figure 4 shows the results for the $P_d$. We see that the $P_d$ is not very sensitive to the smoothing factor for $L \geq 8$. Noting that smaller $L$ means lower complexity, in practice, we can choose a relatively small $L$. However, it is very difficult to choose the best $L$ because it is related to signal property (unknown). Note that energy detection is not affected by $L$.

**(2) Captured DTV signals.** The real DTV signals (field measurements) are collected at Washington D.C., USA. The data rate of the vestigial sideband (VSB) DTV signal is 10.762M samples/sec. The recorded DTV signals were sampled at 21.524476M samples/sec and down converted to a low central IF frequency of 5.381119 MHz (one fourth the sampling rate). The analog-to-digital conversion of the RF signal used a 10-bit or a 12-bit A/D. Each sample was encoded into a 2-byte word (signed int16 with a two's complement format). The multipath channel and SNR of the received signal are unknown. In order to use the signals for simulating the algorithms at very low SNR,

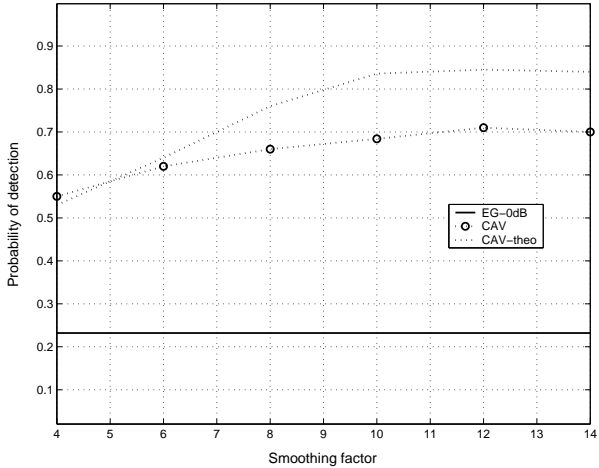

Figure 4: $P_d$ versus smoothing factor for wireless microphone signal: $P_{fa} = 0.01$, $N_s = 50000$, SNR $= -20$dB

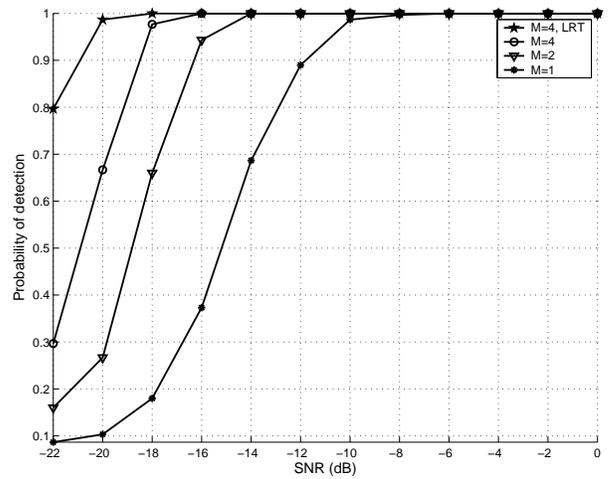

Figure 6: Probability of detection using multiple antennas: $P_{fa} = 0.1$, $N_s = 25000$, one source signal

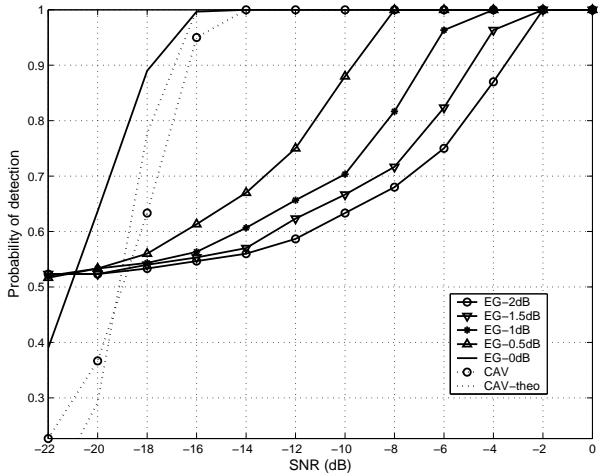

Figure 5: Probability of detection for DTV signal WAS-051/35/01: $N_s = 50000$

we need to add white noises to obtain various SNR levels [23].

Figure 5 gives the simulation results based on the DTV signal file WAS-051/35/01 (receiving antenna is outside and 20.29 miles from the DTV station; antenna height is 30 feet) [21]. The corresponding $P_{fa}$ is shown in Table 1. If noise variance is exactly known ($B = 0$), energy detection is better than the proposed method. However, as discussed in [10, 11, 12], noise uncertainty is always present. Even if noise uncertainty is only 0.5dB, the $P_d$ of energy detection is much worse than that of the proposed method.

In summary, all the simulations above show that the proposed method works well without using information of the signal, the channel and noise power. The energy detection are not reliable (low probability of detection and high probability of false alarm) when there is noise uncertainty.

Thirdly, we simulate the proposed algorithms with multiple antennas/receivers. We consider a system with 4 receiving antennas. Assume that the antennas are well separated (separation larger than half wavelength) such that their channels are independent. This assumption is only for simplicity. In fact, the proposed algorithms perform better if the channels are correlated. Assume that each multipath channel $h_i(k)$ has 5 taps ($N_i = 4$) and all the channel taps are independent with equal power. The channel taps are generated as Gaussian random numbers and different for different Monte Carlo realizations. The source signal $s_0(n)$ are iid and BPSK modulated. The received signal at antenna $i$ is defined in (29). The smoothing factor is $L = 8$ and the number of samples at each antenna is $N_s = 25000$. We fix the $P_{fa} = 0.1$ at all cases. Figure 6 gives the $P_d$ for three cases. From the figure we see that, when only one antenna's signal is used ($M = 1$), the method still works. This verifies our assertion in Section II. B that the method is valid even if the inputs are iid but the channel is dispersive. When 2 ($M = 2$) or 4 ($M = 4$) antennas' signals are combined based on the method in Section II.D, the $P_d$ is much better. The more the antennas we use, the better the $P_d$ is. The optimal LRT [13] detection for $M = 4$ is also included as an upper bound for any detection methods.

To check the performance of the methods for time variant channels, we give a simulation result here. The time variant channel is generated based on the simplified Jake's model. Let $f_d$ be the normalized maximum Doppler frequency (DF). The time variant channel for simulation is defined as

$$h_i(n, k) = \exp\left(j2\pi n \frac{6-k}{5} f_d\right) h_i(k), \ k = 1, \cdots, 5, \quad (84)$$

where $h_i(k)$ is the time invariant channel defined above. For different Doppler frequency from 0 to $10^{-2}$, simulation result is shown in Figure 7 for $M = 2$. For fast time variant channels, the performance of proposed methods

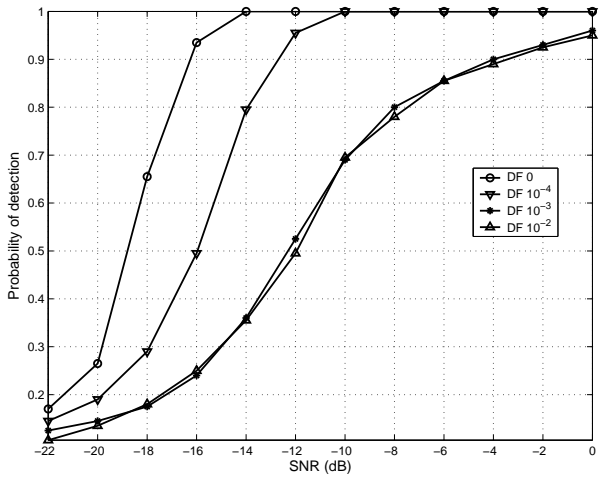

Figure 7: Probability of detection for time variant channels: $M = 2$, $P_{fa} = 0.1$, $N_s = 25000$

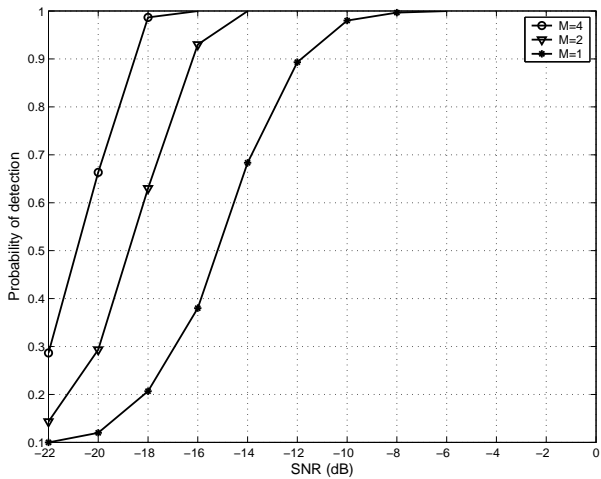

Figure 8: Probability of detection using multiple antennas: $P_{fa} = 0.1$, $N_s = 25000$, three source signals

will degrade.

We also simulate the situation when there are multiple source signals. Figure 8 gives the results for the case of three source signals. Compared to Figure 6, here the results do not change much. Hence, the proposed method is valid when there are multiple source signals.

## 5 Conclusions

In this paper, sensing algorithms based on the sample covariance matrix of the received signal have been proposed. Statistical theories have been used to set the thresholds and obtain the probabilities of detection. The methods can be used for various signal detection applications without knowledge of the signal, the channel and noise power. Simulations based on the narrowband signals, captured DTV signals and multiple antenna signals have been carried out to evaluate the performance of the proposed methods. It is shown that the proposed methods are in general better than the energy detector when noise uncertainty is present. Furthermore, when the received signals are highly correlated, the proposed method is better than the energy detector even the noise power is perfectly known.

## Appendix A

At the receiving end, sometimes the received signal is filtered by a narrowband filter. Therefore, the noise embedded in the received signal is also filtered. Let $\eta(n)$ be the noise samples before the filter, which are assumed to be iid. Let $f(k)$, $k = 0, 1, \cdots, K$, be the filter. After filtering, the noise samples turns to

$$\tilde{\eta}(n) = \sum_{k=0}^{K} f(k)\eta(n-k), \ n = 0, 1, \cdots. \quad (85)$$

Consider $L$ consecutive outputs and define

$$\tilde{\boldsymbol{\eta}}(n) = [\tilde{\eta}(n), \cdots, \tilde{\eta}(n - L + 1)]^T. \quad (86)$$

The statistical covariance matrix of the filtered noise becomes

$$\tilde{\mathbf{R}}_\eta = \mathrm{E}(\tilde{\boldsymbol{\eta}}(n)\tilde{\boldsymbol{\eta}}(n)^T) = \sigma_\eta^2 \mathbf{F}\mathbf{F}^T, \quad (87)$$

where $\mathbf{F}$ is a $L \times (L + K)$ matrix defined as

$$\mathbf{F} = \begin{bmatrix} f(0) & \cdots & f(K-1) & f(K) & \cdots & 0 \\ & \ddots & & & \ddots & \\ 0 & \cdots & f(0) & f(1) & \cdots & f(K) \end{bmatrix}. \quad (88)$$

Let $\mathbf{G} = \mathbf{F}\mathbf{F}^T$. If analog filter or both analog and digital filters are used, the matrix $\mathbf{G}$ should be defined based on those filter properties. Note that $\mathbf{G}$ is a positive definite symmetric matrix. It can be decomposed to

$$\mathbf{G} = \mathbf{Q}^2, \quad (89)$$

where $\mathbf{Q}$ is also a positive definite symmetric matrix. Hence, we can transform the statistical covariance matrix into

$$\mathbf{Q}^{-1}\tilde{\mathbf{R}}_\eta \mathbf{Q}^{-1} = \sigma_\eta^2 \mathbf{I}_L. \quad (90)$$

Note that $\mathbf{Q}$ is only related to the filter. This means that we can always transform the statistical covariance matrix $\mathbf{R}_x$ in (6) (by using a matrix obtained from the filter) such that equation (8) holds when the noise has passed through a narrowband filter. Furthermore, since $\mathbf{Q}$ is not related to signal and noise, we can pre-compute its inverse $\mathbf{Q}^{-1}$ and store it for later usage.